\documentclass[12pt]{article}
\newcommand{\trho}{\tilde{\rho}}

\newcommand{\be}{\begin{equation}}
\newcommand{\ee}{\end{equation}}
{

\begin{document}
\begin{center}
{\large \bf Geometric phases in the simple harmonic and 
      perturbative Mathieu's oscillator systems}
\end{center}
\begin{center}
JeongHyeong Park$^\dag$ and
Dae-Yup Song$^\ddag$
\end{center}
$~~~~~~~~^\dag$Department of Mathematics, Honam University, Kwangju 506-714, Korea\\
$~~~~~~~~~^\ddag$Department of Physics, Sunchon National University, Sunchon 
540-742, Korea

\begin{center}
Short title: Geometric phases \\
Classification numbers: 03.65.Bz  03.65.Ca 03.65.Fd
\end{center}
\begin{abstract}
Geometric phases of simple harmonic oscillator system are studied. Complete sets of 
"eigenfunctions" are constructed, which depend on the way of choosing classical solutions.
For an eigenfunction, two different motions of the probability distribution function 
(pulsation of the width and oscillation of the center) contribute to the geometric 
phase which can be given in terms of the parameters of classical solutions. 
The geometric phase for a general wave function is also given.
If a wave function has a parity under the inversion of space coordinate, then 
the geometric phase can be defined under the evolution of half of the period of 
classical motions.
For the driven case, geometric phases are given in terms of 
Fourier coefficients of the external force.  
The oscillator systems whose classical equation of motion is  Mathieu's equation are 
perturbatively studied, and the first term of nonvanishing geometric phase is calculated. 
\end{abstract}

\section{Introduction}
The harmonic oscillator is a system whose wave functions can be given in terms of 
solutions of classical equation of motion. This fact has been recognized by Lewis 
\cite{Lewis} who found, in an application of asymptotic theory of 
Kruskal \cite{Kruskal},  that there exists a quantum mechanically 
invariant operator which has been used to construct wave functions of (generalized) 
harmonic oscillator systems.
An alternative and simple (at least, conceptually) way to 
construct a complete set of "eigenfunctions" in terms of classical solutions is to use 
the Feynman path integral method \cite{Song1}. Based on these progresses, 
the geometric phases \cite{AA,Berry} for the eigenfunctions of time-periodic harmonic 
oscillator system has been studied \cite{Song2}. The simple harmonic oscillator (SHO) of 
constant mass and constant frequency is a system where any real number can be a period 
of the Hamiltonian and two linearly independent classical solutions are finite all over 
the time. In view of the general analyses  \cite{Song2}, for SHO there is a representation 
where the geometric phase can be defined for an evolution of any real number period. 
For the driven case the phases can not be defined if a particular solution diverges 
\cite{Moore1,Moore2}. 

We note that the geometric phase of SHO has been studied in
\cite{Moore2,Seleznyova}. In considering a spin-1/2 particle in a harmonic 
potential and a nonuniform magnetic field, Pati and Josh consider a wave function 
of SHO whose center oscillates \cite{PJ} and calculate the geometric phase. By adopting 
the fact that a Gaussian wave packet 
could be a wave function of a harmonic oscillator system, geometric phase has been 
calculated \cite{GC} for the wave packet whose width oscillates and relation between 
the phase and Hannay's angle \cite{Hannay} is given for the wave packet.

In this paper, the geometric phases in SHO system will be studied systematically. 
Making use of the general formalism \cite{Song2}, we will construct a complete set of
eigenfunctions and show that, for an eigenfunction, the geometric phase 
comes from two different motions of the probability distribution function: 
the oscillation of the center and the pulsation of the width.  
The driven case will be also considered.
For a harmonic oscillator system whose classical equation of motion is Mathieu's
equation, the geometric phase for an eigenstate will be perturbatively studied in 
a representation which reduces to the "stationary representation" of SHO in a limit.

In the next section, by applying the general formulas \cite{Song2}, the geometric 
phases for the eigenstates of SHO with driving force will given by assuming the 
existence of periodic classical solution. In section 3, the geometric phase of an 
eigenfunction will be interpreted as the contributions from the two different motions. 
In section 4, geometric phase will be calculated for a general wave function of SHO.
In section 5, it will be shown that, if a wave function of SHO has a parity under 
the inversion of space coordinate, the geometric phase of the wave function can 
be defined under the evolution of half of the period of classical motions in SHO.
In section 6, the geometric phase for an eigenstate in Mathieu's oscillator system will 
be perturbatively treated.

\section{General formula}
Classical equation of motion of the SHO with driving force $F(t)$ 
is written as
\be
\ddot{x} + w^2 x = {F(t)\over M}.
\ee
If the two linearly independent homogeneous solutions and a particular solution of (1) 
are denoted as $u(t),v(t)$ and $x_p(t)$, respectively, 
by introducing $\delta(t)$ defined through the relation
\be
\dot{\delta}= {1\over 2}Mw^2x_p-{1\over 2}M\dot{x}_p,
\ee 
the wave functions satisfying Schr\"{o}dinger equation can be written as \cite{Song1,Song3}
\begin{eqnarray}
\psi_m(x,t)&=&<x|m>\cr
  &=&{1\over \sqrt{2^m m!}}({\Omega \over \pi\hbar})^{1\over 4}
     {1\over \sqrt{\rho(t)}}[{u(t)-iv(t) \over \rho(t)}]^{m+{1\over 2}}
     \exp[{i\over\hbar}(M\dot{x}_px+\delta(t) )]
\cr&&\times
     \exp{[{(x-x_p(t))^2\over 2\hbar}(-{\Omega \over \rho^2(t)}
               +i M{\dot{\rho}(t) \over \rho(t)})]}
         H_m(\sqrt{\Omega \over \hbar} {x -x_p(t) \over \rho(t)}),
\end{eqnarray}
where $\rho(t)$ and $\Omega$  are defined as
\be
\rho(t)=\sqrt{u^2(t)+v^2(t)} ~~~~~\Omega=M(u\dot{v} - v\dot{u}).
\ee
With given $u(t),v(t)$ and $x_p(t)$, the wave functions of (3) satisfying 
orthonomality conditions
\be
<n|m>=\int_{-\infty}^\infty \psi_n^*\psi_m dx = \delta_{n,m},
\ee 
form a complete set as can be seen from \cite{Song1}
\be
\sum_m \psi_m(x_b,t_b)\psi_m^*(x_a,t_a)\rightarrow
 \delta(x_b-x_a)
~~~{\rm as}~~~t_b\downarrow t_a.
\ee
Considering this completeness, $\psi_m(x,t)$ will be referred to as an eigenfunction.

The geometric phase for the eigenfunction $\psi_m(x,t)$ is written as \cite{Song2}
\begin{eqnarray}
\gamma_m^F(\tau') &=& {1 \over 2} (m+{1\over 2})\int_0^{\tau'} 
      ({M \dot{\rho}^2 \over \Omega} - {\Omega\over M \rho^2}
         +{Mw^2 \over \Omega}\rho^2)dt + {M \over \hbar}\int_0^{\tau'} \dot{x}_p^2 dt
   \cr
&=&M\int_0^{\tau'} [(m+{1\over 2}){ \dot{\rho}^2 \over \Omega}
+{1\over \hbar}\dot{x}_p^2] dt,
\end{eqnarray}
where $\tau'$ is a period needed for the definition of geometric phase.

Without losing generality, the two linearly independent homogeneous solutions  
$u(t),v(t)$ can be written as \cite{PS}
\be
u(t)= \cos wt ~~~~~~v(t)=C\sin (wt+ \beta),
\ee
where C is a nonzero real number and a real number $\beta$ is not one of  
$(2n+1)\pi/2$ ($n=0, \pm 1, \pm 2,\cdots$). Then $\Omega$ is simply written as
\be
\Omega=MwC \cos\beta .
\ee
For the study of geometric phase, we only consider periodic $F$ satisfying 
$F(t+\tau_f)=F(t)$. For the existence of geometric phase, the $x_p$ should be finite
all over the time which will be assumed from now on.
If there exist two positive integers $N,p$ of no common divisor except 1 such that 
$\tau_0/\tau_f=p/N$, by defining $\tau_0={2\pi / w}$, $N\tau_0$  can be the $\tau'$
in (7). The periodic $F(t)$ can be written as 
\be
F(t)=\sum_{n=-\infty}^\infty f_n e^{in w_f t},
\ee
where 
\be
f_n={1\over \tau_f}\int_0^{\tau_f} F(t) e^{-inw_f t} dt
\ee
with $w_f= 2\pi / \tau_f.$
The $x_p(t)$ finite all over the time can be written as 
\be
x_p=\sum_{n=-\infty}^\infty {f_n \over M(-n^2w_f^2 + w^2)}e^{in w_f t}
    +D\cos(wt+\varphi),      
\ee
where $D$ and $\varphi$ are real numbers. 
If $p=1$, $f_N$ must be zero for the finiteness of $x_p$.
The minimum period of $x_p$ is $N\tau_0$ ($=p\tau_f$) in general so that
\be 
x_p(t+N\tau_0) = x_p(t).
\ee

After some algebra, from (7) one may find the geometric phase of the eigenfunction 
$\psi_n(x,t)$ under $N\tau_0$-evolution as
\begin{eqnarray} 
\gamma_n^F(N\tau_0)
&=&\pi(n+{1\over 2})N[{1-2C\cos\beta +C^2 \over C\cos\beta}]\cr
 && +2\pi{N^3p^2 \over \hbar Mw^3}
    \sum_{n=-\infty}^\infty  {n^2|f_n|^2 \over (p^2n^2 -N^2)^2} 
               + \pi{ MNw \over \hbar} D^2.
\end{eqnarray} 
It should be mentioned that different choice of $(C, \beta, D, \varphi)$
gives different representation (set of eigenfunctions), and thus different geometric phases.
In the next section, an explanation for the fact that the geometric phase does not 
depend on $\varphi$ will be given. 
            
There exist the representations where geometric phase can {\em not} 
be defined even with periodic $x_p$. 
For example, in the representations of $D\neq 1$, if $\tau_f/\tau_0$ is an irrational 
number geometric phases can not be defined under any evolution of finite time.
However, for the SHO with driving force, there is a special representation
of $C=1, \beta=0$ and $D=0$ where geometric phase can be defined for any periodic $x_p$.
In this representation, geometric phase under $\tau_f$-evolution is written as
\be
\gamma_n^F(\tau_f)= {2\pi w_f \over \hbar M}\sum_{n=-\infty}^\infty
       {n^2|f_n|^2 \over (n^2w_f^2 -w^2)^2}.
\ee 

For the SHO without driving force, if we take $C=1$ $\beta=0$ and $x_p=0$, 
the probability distributions of the eigenstates are static, and thus the 
representation of such choice of parameters will be referred to as the stationary 
representation \cite{Song3}. 

In the representation of $C=1, \beta=0$ the geometric 
phase for the driven harmonic oscillator is considered in 
\cite{Seleznyova}, but the given results there do not agree with (14) and (15).

\section{Fictitious particular solution and geometric phase}
From now on we only consider case of $F=0$.

If we take $x_p=0$, 
the geometric phase for $\psi_n$ under the $\tau_0$-evolution is given as 
\be
\gamma_n(\tau_0)
=\pi(n+{1\over 2})[{1-2C\cos\beta +C^2 \over C\cos\beta}].
\ee
Since \cite{Song1}
\be
\int_{-\infty}^\infty\psi_n^*(x,t)x \psi_n(x,t) dx = x_p(t),
\ee
if we take $x_p=0$, the center of the probability distribution function of an eigenfunction 
does not move. In the stationary representation, the probability distribution given from an 
eigenstate does not move as time passes, so that the geometric phase for an eigenstate 
which can be defined for evolution of any period is zero \cite{PS}. 
It has been noted \cite{Song1,Song3} that, except for the stationary representation, 
the widths of probability distributions of the eigenstates pulsate as time passes by, 
which may be the physical origin of the geometric phases given in (16).
The geometric phase for a Gaussian wave packet whose center of probability distribution does 
not move has been considered in \cite{GC}. In order to compare with the the result of 
Ge and Child \cite{GC}, by defining
\[
\alpha(t)={1\over 2\hbar}({\Omega \over \rho^2} -i M {\dot{\rho} \over \rho}),
\]
one may find the relation
\be
-{i\over 2}\int_0^{\tau_0} {\dot{\alpha} \over \alpha+\alpha^*} dt
    = {1-2C\cos\beta +C^2 \over 2 C\cos\beta}\pi
\ee
which gives the geometric phase $\gamma_0(\tau_0)$ in agreement with the result in (16).

Even for the SHO without driving force, one can take the {\em fictitious} particular 
solution as 
\be
x_p=D\cos(wt+\varphi).     
\ee
It may be straightforward, from (3), to find the eigenfunctions for SHO with the $x_p$ in (19)
and check that the eigenfunctions indeed satisfy the Schr\"{o}dinger equation. 
To make the point clear, we only explicitly written down the eigenfunctions of $C=1$ and 
$\beta=0$
\begin{eqnarray}
\psi_n&=&
     \sqrt{\alpha_0 \over 2^n n! \sqrt{\pi}}
     \exp[i\alpha_0^2[-xD\sin(wt+\varphi)+{1\over 4}D^2\sin (2wt+\varphi)]
                     -i(n+{1\over 2})wt]
\cr & &
    \times \exp[-{\alpha_0^2(x-D\cos (wt+\varphi))^2\over 2}]
          H_n(\alpha_0 (x -D\cos (wt+\varphi)) ), 
\end{eqnarray}
where $\alpha_0=\sqrt{Mw/h}.$ From the general formula of (14), one can find that, 
the geometric phase for an eigenfunction $\psi_n$ in (20) under the $\tau_0$-evolution 
is given as $\pi Mw D^2 /\hbar$ $(=\pi\alpha_0^2D^2)$. $\psi_0$ in (20) coincides with
the wave function whose geometric phase is evaluated  in the study of a spin-1/2 
particle in a nonuniform magnetic field \cite{PJ}. The fact that the geometric 
phase does not depend on $\varphi$ is simply understood from that $\varphi$ determines the 
initial point on a cycle while the geometric phase is calculated through the integral 
over a cycle.

This section may be summarized as follows; For an eigenstate in (3) of SHO, 
with the choice of solutions of classical equation of motion in (8,19), the geometric 
phase under $\tau_0$-evolution is written as
\begin{eqnarray} 
\gamma_n(\tau_0)
=\pi(n+{1\over 2})[{1-2C\cos\beta +C^2 \over C\cos\beta}]
               + \pi{ Mw \over \hbar} D^2.
\end{eqnarray} 
The first term in the right-hand side of (21) is from the pulsation of width of
the probability distribution of the eigenstate and the second term is from the oscillation
of the center of the probability distribution.

\section{Geometric phase for a general wave function of SHO}

Due to the completeness, any wave function satisfying Schr\"{o}dinger equation of (driven) 
SHO may be written as a linear combination of eigenfunctions with a specific choice of 
classical solutions:
\be
\Psi=\sum_{n=0}^\infty B_n \psi_{n}(x,t).
\ee
For simplicity, we only consider SHO ($F=0$) system. The classical solutions 
will be chosen as in (8,19). Then one may easily find that every eigenfunction satisfies the 
relation $\psi_n(x, \tau_0+t)= \exp[-i(2n+1)\pi]\psi_n(x,t)$. Therefore, under the
$\tau_0$-evolution, the geometric phase can be defined for any wave function of SHO.

For the explicit evaluation of geometric phase, we need to know the dynamical phase. 
From the formulas in \cite{Song2}, one may find the relation
\begin{eqnarray}
\delta_n(\tau_0)&=& -{1\over \hbar}\int_0^{\tau_0} <n|H| n> dt 
                   = -{1\over \hbar}\int_0^{\tau_0} 
                       <n|({P^2\over 2M}+{1\over 2}Mw^2x^2)| n> dt \cr
 &=& - (n+{1\over 2})\pi{1+C^2 \over C\cos\beta}-\pi \alpha_0^2 D^2.
\end{eqnarray} 
The geometric phase for the wave function in (22) under $\tau_0$-evolution is therefore 
given as
\be
\gamma=\pi[1+\alpha_0^2 D^2
    +\sum_{n=0}^\infty|B_n|^2(n+{1\over 2}){1+C^2 \over C\cos\beta}].
\ee

If $|B_n|$ is given as $|B_n|=\delta_{n,m}$, then one can find that the $\gamma$ given in (24)
is equal to $\gamma_m$ of (21) up to an additive constant $2\pi$.

\section{Geometric phase under the half-period evolution }
Under a time evolution of $\tau_0/2$, any solution of classical equation of motion of SHO 
changes sign with same magnitude.
As noted in \cite{Song2}, the existence of quasiperiodic eigenfunctions depends on the 
periodicity of $\rho(t)$ and $x_p(t)$. Therefore, if we take $x_p=0$ for SHO (that is, in the
absence of fictitious particular solution), the geometric 
phases of the eigenfunctions can be defined under the evolution of half-period $\tau_0/2$.
Indeed, one may find that 
\be
\psi_n(x, \tau_0/2+t)= \exp[-(1/2+n)i\pi]\psi_n(x,t).
\ee
The geometric phase for the eigenfunction $\psi_n|_{x_p=0}$ under the half-period evolution
is given as
\be
\gamma_n(\tau_0/2)={1\over 2}\gamma_n(\tau_0) 
      =(n+{1\over 2})\pi[-1+{1+C^2 \over 2C\cos\beta}].
\ee

Due to the fact that phase is given up to an additive constant $2\pi$, for the wave functions
\[
\Psi_e(x,t)=\sum_{n=0}^\infty B_n^e \psi_{2n}(x,t),
~~~\Psi_o(x,t)=\sum_{n=0}^\infty B_n^o \psi_{2n+1}(x,t),
\]
geometric phases can also be defined under the evolution of the half-period.

Making use of the previous formulas (particularly, (24)), one may easily evaluate the 
geometric phases of $\Psi_e, \Psi_o$. 
In addition, $\Psi_e(x,t),\Psi_o(x,t)$ has the following symmetry
\[ \Psi_e(-x,t)=\Psi_e(x,t)~~~~\Psi_o(-x,t)=-\Psi_o(x,t).\]
Due to the completeness, it may be stated that for a wave having the symmetry 
$\tilde{\Psi}(-x,t)=\pm\tilde{\Psi}(-x,t)$, the geometric phase can be defined under the 
half-period evolution.
 
\section{Perturbative Mathieu's oscillator}

In this section, we will consider the Mathieu's oscillator system described by the 
Hamiltonian 
\be 
H={P^2 \over 2M}+{M\over 2}(a+16\epsilon \cos(2t))x^2,
\ee
where $M,a$ are positive numbers and the real number $\epsilon$  will be assumed small.
The classical equation of motion of the system is given as the Mathieu's equation
\be
\ddot{x}+(a+16\epsilon \cos(2t))x=0.
\ee
 
The Mathieu's equation has long been studied in mathematics and, when $a$ is close to 
$n^2~(n=0,1,2,\cdots)$, some perturbative solutions are known. For example, 
when $a=1-8\epsilon-8\epsilon^2+O(\epsilon^3)$, 
a periodic solution (Mathieu function)  of (28) is known as \cite{WW}
\[
ce_1(t,\epsilon)= \cos t+\sum_{r=1}^\infty\{{2^r\epsilon^r \over (r+1)!r!}
      -{2^{r+1}\epsilon^{r+1} \over (r+1)!(r+1)!} +O(\epsilon^{r+2})\}\cos(2r+1)t.
\]
When $a=1+8\epsilon-8\epsilon^2+O(\epsilon^3)$, there is another series
solution $se_1(t,\epsilon)$ which reduced to $\sin t$ in the limit 
$\epsilon\rightarrow 0$  \cite{WW}.

For the Mathieu's system, if $u(t),v(t)$ are two linearly independent solutions 
of (28), by adopting the definitions in (4), the eigenfunctions and their geometric phases 
are given as in (3) and (7), respectively \cite{Song2}. 
For the calculation of geometric phase, instead of trying to find two linearly 
independent solutions, $\rho(t)$ will be perturbatively evaluated. Similar method 
has been used in \cite{LBF} where, by examining the condition that the perturbative Lewis' 
invariant gives an ellipse in the phase space, the regions of $(a,\epsilon)$ for the 
stable solutions of the Mathieu's equation are given in a reasonably good agreement 
with the exact results in the $\epsilon\rightarrow 0$ limit.  

We rescale the $\rho(t)$ as
\be
\trho=\sqrt{\Omega \over M}\rho.
\ee
Then one may find \cite{Song1,Lewis} that the $\trho$ satisfies the following 
differential equation
\be
\ddot{\trho}+(a+16\epsilon \cos(2t))\trho -{1\over \trho^3}=0.
\ee
The geometric phase for an eigenfunction $\psi_n(x,t)$ may then be written as
\be
\gamma_n^M=(n+{1\over 2})\int_0^\pi \dot{\trho}^2 dt,
\ee
if $\trho$ is periodic under $\pi$-evolution.

We assume the perturbative expansion of $\trho$ for small $\epsilon$ as
\be
\trho=\trho_0+\epsilon\trho_1+\epsilon^2\trho_2+\epsilon^3\trho_3+O(\epsilon^4).
\ee
From (30), one may find the differential equation for the $\trho_0$  
\be
\ddot{\trho}_0+ a\trho_0-{1\over \trho_0^3}=0.
\ee
We will assume that $\trho_0$ is constant, which gives 
\be
\trho_0=a^{-1/4}.
\ee
The above assumption amounts to that we only consider the representation(s) of Mathieu's
oscillator system which reduces to the stationary representation of SHO in the 
$\epsilon\rightarrow 0$ limit. 
Such assumption may be natural except for the cases $a=(n)^2$ $(n=1,2,\cdots)$, 
since we should find, for the geometric phase, the representation of quasiperiodic 
eigenfunctions under the evolution of $\pi$ in the Mathieu's system (which reduces to 
the SHO in the $\epsilon\rightarrow 0$ limit), while for $a\neq n^2$ \cite{PS} the 
stationary representation is the unique one of quasiperiodic eigenfunctions under the 
$\pi$-evolution in the SHO system of $\epsilon=0$. 

By comparing terms of same order in (30), we find the recursive equations:
\begin{eqnarray}
\ddot{\trho}_1+4a\trho_1 =-16{\cos 2t \over a^{1/4}},~~~
\ddot{\trho}_2+4a\trho_2 =-16\trho_1 \cos 2t + 6 a^{5/4}\rho_1^2,~~~
\cdots .
\end{eqnarray}
The solutions are given as
\begin{eqnarray}
\trho_1 &=&-{4\cos 2t \over a^{1/4} (a-1)}, \\
\trho_2 &=& {4(5a-2) \over a^{5/4}(a-1)^2} 
           +{4(5a-2) \over a^{1/4}(a-1)^2(a-4)}\cos 4t,~~~\cdots .
\end{eqnarray}
In obtaining  (36,37), we require that $\trho_1,\trho_2\cdots$ are periodic functions of
$\pi$ for general $a$. 
From (31), the geometric phase for the eigenfunction $\psi_n$ in the representation 
mentioned above is given as
\be
\gamma_n^M=\epsilon^2 {8(n+{1\over 2})\pi \over \sqrt{a}(a-1)^2}+O(\epsilon^4).
\ee
In evaluating (38), $\trho_2$ has been used to show the absence of  a term of
the order of $\epsilon^3$ in $\gamma_n^M$.

The expressions of $\trho_1,\trho_2$ in (36,37) indicate that this perturbative method is
not valid when $a=1,4$ (presumably, also when $a=n^2~~(n=3,4\cdots))$. In \cite{Song2},
it has been proven that geometric phase can not be defined for a harmonic oscillator 
system if one of the classical solutions diverges as time goes to infinity. 
Indeed, it has been 
numerically proved that for the cases $a=1$ or $a=4$ with nonzero $\epsilon$, one of the 
solution diverges as time goes to infinite \cite{AS} (see also \cite{Magnus,Song2}).
The numerical study \cite{AS} also indicates the existence of the representation
(the existence of {\em real} characteristic exponent) 
where the geometric phase can be defined for small $\epsilon$ and $a\neq n^2$ (see
\cite{Song2}).

\section{Summary and discussions}

The general formalism for the geometric phases of harmonic oscillators \cite{Song2} has been 
explicitly applied for (driven) SHO. By calculating the geometric 
phases for the eigenfunctions, and relying on the completeness of a set of eigenfunctions, 
geometric phase for a general wave function of SHO is given.

The explicit study on the SHO system suggests a new interesting feature on the geometric 
phases of harmonic oscillator \cite{Song2} whose Hamiltonian is periodic with 
period $\tau_H$; For the case of two linearly independent 
periodic homogeneous solutions of period $\tau_H$ or $2\tau_H$, we can think of taking 
fictitious particular solution even for the undriven oscillator, which would give 
different geometric phases without changing the period needed for 
the definition of geometric phase. However, for the case where one of the homogeneous 
solution is {\em not} periodic with period $\tau_H$ or $2\tau_H$, 
if we take fictitious particular solution for the $F=0$ oscillator system, 
the period  needed for the definition of geometric phase should be changed or 
in some cases the geometric phase can not 
be defined.

The geometric phase of an eigenfunction of SHO is interpreted as the contributions from the 
two different motions of probability distribution (pulsation of width and oscillation 
of the center). It has been shown in \cite{Seleznyova}, the eigenfunction whose center 
oscillate is related to the coherent state constructed by applying displacement operator 
to the ground state of the stationary representation. As suggested in \cite{Song3}, 
the eigenfunctions of harmonic oscillator is closely related to the generalized coherent 
states, and the eigenfunctions of probability distribution function whose width 
pulsates would be related to the generalized coherent states constructed by applying 
the squeezing operators to the ground state of the stationary representation  \cite{Coh}.  
The geometric phases has been studied for a generalized coherent states in \cite{co}.
It would be interesting to compare our results with those for coherent states.
The contribution to the geometric phase from the motion of the center of probability
distribution ($\dot{x}_p$) for the driven case is $O(1/\hbar)$ in a general 
harmonic oscillator. For the $F=0$ case, the contribution from the oscillation of 
the center which is from taking fictitious particular solution 
is also  $O(1/\hbar)$.

We also studied the geometric phase of perturbative Mathieu's oscillator system of 
small $\epsilon$. The representation in which we calculate the geometric phase must 
be the only one representation of the system where the geometric phase can be defined 
under $\pi$-evolution,
since the stationary representation is the only one representation of the corresponding
SHO ($\epsilon=0$) system with quasiperiodic eigenfunctions except for the cases 
$a=(n)^2$ $(n=1,2,\cdots)$. 

As a final remark, we note that geometric phase has been studied in the finite
dimensional Hilbert space of SHO \cite{PL} constructed from bra and ket vectors
in the stationary representation, which is based on the Pegg-Barnett formalism 
\cite{PB} of finite dimensional state space (see, also \cite{BWKL}).
It would be interesting to study the geometric phases for the finite dimensional 
Hilbert space in a representation other than the stationary one.

\bigskip\bigskip
\noindent
{\bf Acknowledgments} \newline
JHP wishes to acknowledge the 
partial financial support of the Korea Research Foundation made in the program, 
year of 1998. DYS acknowledges the partial financial support
from Sunchon National University (Non-Directed Research Fund).

\end{document}